\documentstyle[12pt]{article}
\begin{document}

\begin{center}
\Large {\bf Towards a more exact value of deflection of light due
to static gravitational mass}

\vspace{0.5cm}

\large {\bf  {A. K. Sen}}

 Department of Physics, Assam University, Silchar 788011, India

 Email: asokesen@bsnl.com
\end{center}

\bigskip
\bigskip

{\bf Abstract} The deflection of a ray of light passing close to a
gravitational mass, is generally calculated from the null geodesic
which the light ray ( photon)  follows. However, there is an
alternate approach, where the effect of gravitation on the ray of
light is estimated by considering the ray to be passing through a
\textit{material medium}.  Calculations have been done in this
paper, following the later approach,  to estimate the amount of
deflection due to a static non-rotating mass. The refractive index
of such a material medium, has been  calculated in a more rigorous
manner in the present work and the final expression for the amount
of  deflection calculated here is claimed to be more accurate than
all other expressions derived so far using \textit{material
medium} approach. Based on this expression, the amount of
deflection for a sun grazing ray  has been also calculated.

\section{Introduction}

The gravitational deflection of light is one of the important
predictions of the General Theory of Relativity (GTR) proposed by
Einstein, which plays a key role in understanding problems related
to Astronomy, Cosmology, Gravitational Physics and other related
branches.

Newton's theory of universal gravitation had already predicted
that the path of any material particle moving at a finite speed is
affected by the pull of gravity. By the late 18th century, it was
possible to apply Newton's law to compute the deflection of light
by gravity. Cavendish commented briefly on the gravitational
deflection of light in the late 1700s and Soldner gave a detailed
derivation in 1801.

The idea of bending of light was revived by Einstein in 1911 and
the quantitative prediction for the amount of deflection of light
passing near a large mass (M) was identical to the old Newtonian
prediction, $d = 2GM/(c^2 r_{\bigodot})$, where $r_{\bigodot}$ is
the  closet distance of approach  and in this case approximately
the solar radius. It wasn't until late in 1915, as Einstein
completed the general theory, he realized his earlier prediction
was incorrect and the angular deflection should actually be twice
the size he predicted in 1911. This was subsequently confirmed by
Eddington in 1919 through an experiment performed during the solar
eclipse.

The exact amount of deflection for a ray of light passing  close
to a gravitational mass can be worked out from the  null geodesic,
which a ray of light follows [1,2,3].

The deflection of a light ray passing close to a gravitational
mass can be alternately calculated by following an approach, where
the effect of gravitation on the light ray is estimated by
considering the  light ray to be passing through a material medium
with a value of refractive index decided by the value of
gravitational field [4].

The concept of this \textit{equivalent material medium} was
discussed by Balazs [5] as early as in 1958,  to calculate the
effect of a rotating body, on the polarization of an
electromagnetic wave passing close to it. Plebanski [6] had also
utilized this concept in 1960, to study the scattering of a plane
electromagnetic wave by gravitational field, where the author
mentioned that this concept of {\it equivalent material medium}
was first pointed out by Tamm [7] in 1924. A general procedure for
utilizing this concept, for deflection calculation has been worked
out by Felice [8]. Later this concept was also used by Mashoon
[9,10], to calculate the deflection and polarization due to the
Schwarzschild and Kerr black holes. Fischbach and Freeman [11],
derived the effective refractive index of the material medium and
calculated the second order contribution to the gravitational
deflection. In a similar way Sereno [12] has used this idea, for
gravitational lensing calculation by drawing the trajectory of the
ray by Fermat's principle. More recently Ye and Lin [13],
emphasized the simplicity of this approach and calculated the
gravitational time delay and the effect of lensing.

On the other hand, the calculation of higher order deflection
terms, due to Schwarzschild Black hole, from the null geodesic,
has been performed recently by Iyer and Petters  ( [14]  and
references their in). Using null geodesics, gravitational lensing
calculations have been done by a number of authors in past
[15,16].

With the above background, in the present work, we follow the {\it
material medium} approach, to calculate a more accurate value for
the deflection term due to a non-rotating sphere ( Schwarzschild
geometry ).  It is claimed that the present calculated value will
be more accurate than all other values calculated in past, using
{\it material medium } concept.

\section{The effective refractive index and the trajectory of light ray}

As discussed earlier, the gravitational field influences the
propagation of electromagnetic radiation by imparting to the space
an effective index of refraction $n(r)[4]$.

For a static and spherically symmetric gravitational field, the
solution of Einstein's Field Equation was  given by K.
Schwarzschild in 1961, which is as follows[4]:

\begin{equation}
ds^2= (1-\frac{r_g}r)c^2 dt^2 - r^2 ( sin^2 \theta d \phi^2 +d
\theta ^2) -\frac {dr^2}{(1-\frac {r_g}{r})}
\end{equation}

where $r_g=\frac {2km}{c^2}$ called Schwarzschild Radius, which
completely defines the gravitational field in vacuum produced by
any centrally-symmetric distribution of masses. The above equation
can be expressed in an isotropic form by introducing a new radius
co-ordinate ( $\rho$ ) with the following transformation equation
[4]

\begin{equation}
\rho= \frac{1}{2} [(r-\frac{r_g}2)+ r^{1/2} (r-r_g)^{1/2}]
\end{equation}

\centerline {OR}
\begin{equation}
r = \rho (1 + \frac {r_g} {4 \rho})^2
\end{equation}

The resulting isotropic form of Schwarzschild equation will be
now:

\begin{equation}
ds^2= (\frac{1-r_g/(4\rho)}{1+r_g/(4\rho)})^2 c^2 dt^2 - (1+
\frac{r_g}{4 \rho})^4(d\rho ^2 + \rho ^2 ( sin ^2 \theta  d \phi ^
2 + d \theta ^ 2))
\end{equation}

Now in spherical co-ordinate system the quantity $(d\rho ^2 + \rho
^2 ( sin ^2 \theta  d \phi ^ 2 + d \theta ^ 2))$ has the dimension
of square of infinitesimal length vector $d \overrightarrow{
\rho}$.

By setting $ds=0$,  the  velocity of light can be identified from
the expression of the form $ds^2 = f (\rho)dt ^2 - d
\overrightarrow{ \rho} ^2 $, as $v(\rho)= \sqrt {f(\rho )}$.
Therefore the velocity of light in the present case (
characterized by Schwarzschild radius $r_g$ ) can be expressed as
:

\begin{equation}
v( \rho ) = \frac {(1-\frac{r_g}{4\rho })  c }{(1+ \frac {r_g}{4
\rho})^3}
\end{equation}

But this above expression of velocity of light is in the unit of
length $\rho$ per unit time. We therefore write

\begin{eqnarray}
v(r)=&& v(\rho) \frac {dr}{d\rho}\nonumber\\
&&= v(\rho) [(1+ \frac {r_g}{4 \rho})^ 2 - \frac{r_g}{2 \rho}(1+
\frac {r_g}{4\rho})]\nonumber\\
&&= (\frac{r_g-4\rho}{r_g+4\rho})^2 c
\end{eqnarray}

 Substituting the value of $\rho$ from Eqn (2) in Eqn.(6),  we get:

\begin{eqnarray}
v(r) =&&(\frac{r_g/2- 2\rho}{r_g/2+2\rho})^2 c\nonumber\\
=&&(\frac{r_g/2-((r-\frac{r_g}2)+ r^{1/2} (r-r_g)^{1/2})}{r_g/2+ ((r-\frac{r_g}2)+ r^{1/2} (r-r_g)^{1/2})})^2 c\nonumber\\
=&&(\frac{r_g-r- r^{1/2} (r-r_g)^{1/2}}{r+ r^{1/2} (r-r_g)^{1/2}})^2 c\nonumber\\
=&& \frac {c(r-r_g)}{r}
\end{eqnarray}

Therefore the refractive index $n(r)$ at a  point  with spherical
polar co-ordinate $(r)$, can be expressed by the relation:

\begin{equation}
n(r)=\frac{c}{v(r)}=\frac{r}{r-r_g}
\end{equation}

At this stage the entire problem, can become a problem of
geometrical optics, where we have to find the trajectory of a
light ray travelling in a medium, whose refractive index has
spherical symmetry. The trajectory of the light ray and the center
of mass ( source of gravitational potential) will together define
a plane. The equation of such a ray in plane polar co-ordinate
system $(r,\theta)$ can be written as [17]:

\begin{equation}
\theta = A. \int^{\infty}_{r_{\bigodot}}{\frac {dr}{r
\sqrt{n^2r^2-A^2}}}
\end{equation}

The trajectory is such that $n(r).d$ always remains a $constant$,
where $d$ is the perpendicular distance between the trajectory of
the light ray from the origin and the \textit{constant} is taken
here as $A$ [17]. In our present problem the light is coming from
infinity ( $r=- \infty $) and it is approaching the gravitational
mass, which is placed at the origin and  characterized by
Schwarzschild radius $r_g$. The closest distance of approach,  for
the approaching ray is $b$ and the ray goes to $r=\infty $, after
undergoing certain amount of deflection ($\triangle \phi$).

Here, the parameter $b$ can be replaced by solar radius
$r_{\bigodot}$. When the light ray passes through the  closest
distance of approach (ie $r=b$ or $r_{\bigodot}$), the tangent to
the trajectory becomes perpendicular to the vector
${\overrightarrow r}$ ( which is $ \overrightarrow r_{\bigodot}$).
Therefore, we can write $A= n(r_{\bigodot})r_{\bigodot}$. The
trajectory of the light ray had been already constructed before
like this, by Ye and Lin [13] and the value of deflection
($\triangle \phi$), can be written as :

\begin{equation}
\triangle \phi = 2 \int^{\infty}_{r_{\bigodot}} { \frac {dr}{r
\sqrt{(\frac{n(r).r}{n(r_{\bigodot}).r_{\bigodot}})^2-1}}} - \pi
\end{equation}

However, Ye and Lin [13], had in our opinion  used a value of
refractive index $n(r)$ which was approximated and  somewhat ad
hoc. Fischbach and Freeman [11] also in their attempt to calculate
a more accurate value of deflection, considered terms only up to
second order in the expression for refractive index. However, in
our attempt to do so we shall avoid making any such approximation
in the following. We denote the above integral in Eqn. (10) by $I$
and write

\begin{eqnarray}
I=&& \int ^{\infty}_{r_{\bigodot}}{ \frac {dr}{r \sqrt{(\frac{n(r).r}{n(r_{\bigodot}).r_{\bigodot}})^2-1}}}\nonumber\\
&&= n(r_{\bigodot})r_{\bigodot}\int^{\infty}_{r_{\bigodot}} { \frac {dr}{r \sqrt{(n(r).r)^2-(n(r_{\bigodot}).r_{\bigodot})^2}}}\nonumber\\
&&= n(r_{\bigodot})r_{\bigodot}\int ^{\infty}_{r_{\bigodot}}{
\frac {dr}{r \sqrt{ \frac{r^4}{(r-r_g)^2}-\frac{r_{\bigodot}
^4}{(r_{\bigodot}-r_g)^2}}}}\nonumber\\
 &&= n(r_{\bigodot})r_{\bigodot}\int^{\infty}_{r_{\bigodot}}{
\frac {dr}{r^2 \sqrt{
\frac{1}{(1-\frac{r_g}{r})^2}-\frac{r_{\bigodot} ^2 r^{-2}
}{(1-\frac{r_g}{r_{\bigodot}})^2}}}}
\end{eqnarray}

Now we change the variable to $x=\frac{r_g}{r}$ and introduce a
quantity $a=\frac{r_g}{r_{\bigodot}}$. We also denote
$n(r_{\bigodot})$ by $n_{\bigodot}$. Accordingly we write:

\begin{eqnarray}
I=&& n_{\bigodot}r_{\bigodot}\int^{0}_{a} { \frac {-x^{-2}r_g dx}{r^2 \sqrt{\frac{1}{(1-x)^2}-\frac{x^2}{(a(1-a))^2}}}}\nonumber\\
&&= n_{\bigodot}r_{\bigodot}\int^{0}_{a} { \frac {-x^{-2}r_g dx}{x r^2 \sqrt{\frac{1}{(x(1-x))^2}-\frac{1}{(a(1-a))^2}}}}\nonumber\\
&&= \frac{n_{\bigodot}r_{\bigodot}}{r_g}\int^{a}_{0} { \frac
{dx}{x
\sqrt{\frac{1}{(x(1-x))^2}-\frac{1}{(a(1-a))^2}}}}\nonumber\\
&&= \frac{n_{\bigodot}r_{\bigodot}}{r_g}\int^{a}_{0} { \frac
{(1-x)dx}{ \sqrt{1-\frac{(x(1-x))^2}{(a(1-a))^2}}}}
\end{eqnarray}

For our convenience we can denote the quantity $1/(a(1-a))$ by
$D$. This also implies

\begin{equation}
D =\frac{r_{\bigodot} ^2}{r_g( r_{\bigodot} - r_g)}
\end{equation}

However, the Integral $I$ can not be solved as a standard integral
at this stage.  We split the above Integral, as a sum of two
Integrals and proceed as follows:

\begin{eqnarray}
I=&& (\frac {n_{\bigodot}r_{\bigodot}}{r_g})[\int^{a}_{0} { \frac {(1-2x) dx}{\sqrt{ 1-D^2x^2(1-x)^2}}} +  \int^{a}_{0} { \frac {x dx}{\sqrt{ 1-D^2x^2(1-x)^2}}}]\nonumber\\
&&=(\frac {n_{\bigodot}r_{\bigodot}}{r_g})\int^{a}_{0} { \frac {(1-2x) dx}{\sqrt{ 1-D^2x^2(1-x)^2}}} + (\frac {n_{\bigodot}r_{\bigodot}}{r_g}) \int^{a}_{0} { \frac {x dx}{\sqrt{ 1-D^2x^2(1-x)^2}}}\nonumber\\
&&=( \frac{n_{\bigodot}r_{\bigodot}}{r_g}) I_1 + (\frac
{n_{\bigodot}r_{\bigodot}}{r_g}) I_2
\end{eqnarray}

where $I_1$ and $I_2$ are used to denote the above two integrals.
Now we can identify

$$\frac{n_{\bigodot} r_{\bigodot}}{r_g} = \frac{1}{1-a}. \frac{1}{a}= \frac {1}{a(1-a)}=D$$
Changing the variable from $x$ to $y=D x (1-x)$, we can write
$D(1-2x)dx=dy$. Accordingly the upper and lower limits $x= 0$ and
$x = a$ change to $y=0$ and
$y=Da(1-\frac{r_g}{r_{\bigodot}})=\frac {1}{a(1-a)}a(1-a)=1$.
Therefore for the first part in Eqn (14) we can write :

\begin{eqnarray}
(\frac{n_{\bigodot}r_{\bigodot}}{r_g}) I_1=&&  \int^{a}_{0} { \frac {D(1-2x) dx}{\sqrt{ 1-D^2x^2(1-x)^2}}} \nonumber\\
&&=\int^{1}_{0} \frac {dy}{\sqrt{1-y^2}}\nonumber\\
&&=[sin^{-1}y]^1_0\nonumber\\
&&=\pi/2
\end{eqnarray}

Therefore, from Eqn (10), one may write the amount of deflection
as:

\begin{eqnarray}
\triangle \phi =&& 2 \int^{\infty}_{r_{\bigodot}} { \frac {dr}{r
\sqrt{(\frac{n(r).r}{n(r_{\bigodot}).r_{\bigodot}})^2-1}}} - \pi \nonumber\\
&&=2 (\frac {n_{\bigodot}r_{\bigodot}}{r_g})I_1 + 2(\frac
{n_{\bigodot}r_{\bigodot}}{r_g}) I_2 - \pi \nonumber\\
&&=\pi + 2(\frac {n_{\bigodot}r_{\bigodot}}{r_g})I_2 - \pi \nonumber\\
&&=(\frac {2 n_{\bigodot}r_{\bigodot}}{r_g})\int^{a}_{0}{\frac {x
dx}{\sqrt{ 1-D^2x^2(1-x)^2}}}
\end{eqnarray}

Thus the  gravitational bending for a ray of light grazing the
static gravitational mass ( with Schwarzschild radius $r_g$ ) with
the closest distance of approach  $r_{\bigodot}$ can be expressed
as:

\begin{equation}
\triangle \phi = 2 D \int^{a}_{0} { \frac {x dx}{\sqrt{
1-D^2x^2(1-x)^2}}}
\end{equation}

The above expression for gravitational deflection has been
obtained from the Schwarzschild Equation ( Eqn(1)), with out
applying any  approximation at any stage. Owing to this, it is
claimed that this expression of bending is  more exact as compared
to all other expressions derived till today, using {\it equivalent
material medium} concept. However, the integration of the quantity
in Eqn (17)), involves some complicated algebraic expressions
containing Elliptical functions. Using Mathematica, we obtain the
following expression after integration :

\begin{eqnarray}
\int {\frac{x dx}{\sqrt{1-D^2 x^2 (1-x)^2}}}=2\frac{(\sqrt{D}+\sqrt{D-4})E-(2\sqrt{D-4})F}{D(\sqrt{D+4}-\sqrt{D-4})}\nonumber\\
\end{eqnarray}

where $E\equiv E(p,q^2)$ is the Elliptic Integral of first kind
and $F\equiv F(-q,p,q^2)$ is Incomplete Elliptic Integral of Third
kind. The arguments p,$q^2$,-q,p,$q^2$ are expressed by the
following mathematical relations:

\begin{eqnarray}
p=&& \arcsin {\sqrt{\frac
{(\sqrt{D-4}-\sqrt{D+4})(\sqrt{D-4}+(2x-1)\sqrt{D})}
{(\sqrt{D-4}+\sqrt{D+4})(\sqrt{D-4}-(2x-1)\sqrt{D} )}}} \nonumber\\
\end{eqnarray}

\begin{eqnarray}
q=&&\frac{(\sqrt{D-4}+\sqrt{D+4})}{(\sqrt{D-4}-\sqrt{D+4})} \nonumber\\
\end{eqnarray}

Finally we can write the expression for gravitational deflection
($\triangle \phi$) of  the light ray, due to a  static mass
($r_g$) with the closest distance of approach $r_{\bigodot}$ as :

\begin{equation}
\triangle \phi
=4\left\{\frac{(\sqrt{D}+\sqrt{D-4})E-(2\sqrt{D-4})F}{(\sqrt{D+4}-\sqrt{D-4})}\right\}^{x=a}_{x=0}
\end{equation}

where the value of $D$ is given by Eqn.(13) as
$D=\frac{r_{\bigodot}^2}{r_g(r_{\bigodot}-r_g)}$ and
$a=r_g/r_{\bigodot}$. Eqn. (21) is a general expression for
bending of light, where $r_{\bigodot}$ can be replaced by the
closest distance of approach of the light ray. This mathematical
expression for deflection, derived here is claimed to be more
accurate than all other expressions derived so far using
\textit{material medium} approach and it is equally valid for
strong field. For a Sun grazing ray, we can take the closest
distance of approach as equal to solar radius which is
$r_{\bigodot}= 695,500$ km and Schwarzschild radius corresponding
to the mass of Sun as $r_g= 3$ km. We, therefore, get
$a=(r_g/r_{\bigodot})=1/231,833$ and $D=231,834$. Finally, we get
a value of $\triangle \phi = 8.62690 E10^{-6}$ radians or 1.77943
$ arc$ $sec$. This value of gravitational deflection suffered by a
Sun grazing ray, is claimed to be more accurate than all  other
values obtained in past.

{\bf Acknowledgments:} Some of the calculations reported in this
paper were done using Mathematica, at IUCAA, Pune, India.  I
sincerely acknowledge this help and support.

\newpage 

\bf{References}

[1] C W Misner, K S Thorne and J A Wheeler,
\texttt{Graviation},(W. H. Freeman and Company, Newyork 1972)

[2] S. Weinberg, \textit{Principles and Applications of the
General Theory of  Relativity},(John Wiley \& Sons Inc. 1972)

[3] P. Schneider, P. Ehlers, E. E. Falco, \textit{Gravitational
 Lenses}, (Springer, Berlin 1999)

[4] {\it The Classical Theory of Fields volume 2; } J. L. D.
Landau and E. M. Lifshitz (1st edition Pergamon Press, 1951; 4 th
edition' Butterworth-Heinemann 1980)

[5] N. L. Balazs , Phys. Rev.\textbf{110}, No. 1, 236 (1958)

[6] J. Plebanski, Phys. Rev. \textbf{118}, No.5, 1396 (1960)

[7] J. E. Tamm, J. Russ. Phys.-Chem. Soc. \textbf{56},2-3, 284
(1924)

[8]F de Felice, Gen. Relativ. Gravit. \textbf{2}, No. 4, 347
(1971)

[9] B Mashhoon, Phys. Rev. D  \textbf{7},  No. 10, 2807 (1973)

[10]B Mashhoon, Phys. Rev. D  \textbf{11},  No. 10, 2679 (1975)

[11] E. Fischbach and B. S. Freeman, Phys. Rev. D \textbf{67},
(1980)

[12]M Sereno, Phys. Rev. D \textbf{67}, (2003)

[13] X-H. Ye and  Q. Lin , Journal Mod. Opt.  \textbf{55}, Issue
7, 1119 (2007)

[14] S. V. Iyer and A. O. Petters, Gen. Relativ. Gravit.
\textbf{39}, 1563 (2007)

[15] K. S. Virbhadra and George  F. R. Ellis, Phys. Rev. D
\textbf{62}, 084003 (2000)

[16] S. Frittelli, T.P. Kiling, and E.T. Newman, Phys. Rev. D.
\textbf{62}, 064021 (2000)

[17]M. Born and E wolf 1947, {\it Principles of Optics} (7th
Edition, Cambridge University Press, Cambridge,1999)p121

\end{document}